\documentclass[fleqn,twoside]{article}
\usepackage{espcrc2}
\usepackage{epsfig}
\usepackage{amssymb}
\usepackage{amsbsy}
\usepackage{amsfonts}
\usepackage{graphics}
\usepackage{latexsym}
\usepackage{epsf}

\newcommand{\fm}{\mbox{fm}}
\newcommand{\sgn}{\mbox{sgn}}
\newcommand{\MeV}{\mbox{MeV}}
\newcommand{\GeV}{\mbox{GeV}}
\newcommand{\eq}{\begin{equation}}
\newcommand{\ee}{\end{equation}}
\newcommand{\ea}{\begin{eqnarray}}
\newcommand{\eea}{\end{eqnarray}}
\usepackage{graphicx}
\usepackage[figuresright]{rotating}

\title{\vspace{-2.25cm}
       {\normalsize DESY 04--191}    \\[-0.2cm]
       {\normalsize Edinburgh 2004/23} \\[-0.2cm]
       {\normalsize LTH 636} \\[-0.2cm]
       {\normalsize LU-ITP 2004/037} \\[0.50cm]
A lattice determination of $g_A$ and $\langle x\rangle$ from overlap 
fermions\thanks{Talk presented by T. Streuer at Lattice 2004.}}

\author{
M. G\"urtler\address{John von Neumann-Institut f\"ur Computing 
NIC, Deutsches Elektronen-Synchrotron DESY,\\ 15738 Zeuthen, Germany},
R. Horsley\address{School of Physics, University of Edinburgh, 
Edinburgh EH9 3JZ, UK},
V. Linke\address{Institut f\"ur Theoretische Physik, Freie 
Universit\"at Berlin, 14196 Berlin, Germany},
H. Perlt\address{Institut f\"ur Theoretische Physik, Universit\"at 
Regensburg, 93040 Regensburg, Germany},
P.E.L. Rakow\address{Theoretical Physics Division, Department of 
Mathematical Sciences, University of Liverpool,\\ Liverpool L69 3BX, UK},
G. Schierholz$^{\rm a,}$\address{Deutsches Elektronen-Synchrotron  
DESY, 22603 Hamburg, Germany},
A. Schiller\address{Institut f\"ur Theoretische Physik, Universit\"at 
Leipzig, 04109 Leipzig, Germany} and
T.~Streuer$^{\rm a,c}$}

\begin{document}

\begin{abstract}
We present results for the nucleon's axial charge $g_A$ and the first
moment $\langle x \rangle$ of the unpolarized parton distribution
function from a simulation of
quenched overlap fermions.
\end{abstract}

\maketitle

\section{INTRODUCTION}

The axial charge $g_A$ of the nucleon
describes the beta decay of the neutron. It can be defined by
\eq
2s_\mu g_A=\langle p(\vec{p}),\vec{s}|(A^{(u)}_\mu-A^{(d)}_\mu)|p(\vec{p}),\vec{s}\rangle
\label{eq_ga2}
\ee
where $\vec{s}$ is the spin vector of the proton and
$A^{(q)}_\mu~=~\bar{q}\gamma_\mu\gamma_5 q$.
Thus, $g_A$ can be computed on the lattice from a flavour non-singlet
proton matrix element.
There have been lattice determinations of $g_A$ in the past;
most of them used Wilson-like fermions, which makes it hard to go to
small quark masses. For recent results from Wilson fermions see e.~g. \cite{goeckeler2}.

For the first moment of the unpolarised parton distribution $\langle x\rangle$
, which measures the fractional momentum carried by the quarks, the comparison of lattice and phenomenological data has been problematic \cite{bakeyev1}. To settle the problem, one needs to do simulations at smaller quark masses.

Here, we present results for these two quantities obtained with quenched
overlap fermions.\\

\section{SIMULATION DETAILS}

We use the overlap operator given by

\eq
D=\rho\big(1+\frac{m_q}{2\rho}+(1-\frac{m_q}{2\rho})
\gamma_5 \sgn(H_W(-\rho))  \big)\, 
\label{eq_D}
\ee
where $H_W(-\rho)=\gamma_5 (D_W-\rho)$, $D_W$ being the Wilson Dirac operator.
We approximate the
sign function appearing in (\ref{eq_D}) by minmax polynomials \cite{giusti1}.
For the gauge part we chose the L\"uscher-Weisz action \cite{luscher1}

\begin{figure}[t]
\begin{center}
\epsfig{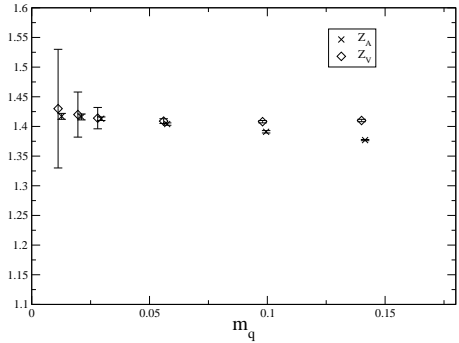}
\vspace*{-0.5cm}
\caption{Comparison between $Z_V$ obtained from the nucleon matrix element
 (eq. (\ref{eq_ZA})) and $Z_A$ obtained from the Ward identity
 (eq. (\ref{eq_ZV})).}
\label{plot_ZAZV}
\end{center}
\vspace*{-1cm}
\end{figure}

\begin{eqnarray}
S[U]&\!\!\!=\!\!\!&\frac{6}{g^2}\Big\{ c_0\!\!\!\! 
\sum_{\rm plaquette}\frac{1}{3}\,
\mbox{Re}\, 
\mbox{Tr}\, [1-U_{\rm plaquette}] \nonumber \\
&\!\!\!+\!\!\!& c_1\!\!\!\! \sum_{\rm rectangle}\frac{1}{3}\, \mbox{Re}\, 
\mbox{Tr}\, [1-U_{\rm rectangle}] \\
&\!\!\!+\!\!\!& c_2\!\!\!\!\!\!\!\! \sum_{\rm parallelogram}\frac{1}{3}\, 
\mbox{Re}\, \mbox{Tr}\, [1-U_{\rm parallelogram}] \Big\} \nonumber
\label{eq_LW}
\end{eqnarray}
\vspace*{-0.4cm}

\noindent
with coefficients $c_1$, $c_2$ ($c_0 + 8 c_1 + 8 c_2 = 1$) taken from tadpole 
improved perturbation theory~\cite{gattringer1}.
We ran our computations at $\beta=8.45$, corresponding to $a=0.095\fm$, at
two volumes, $V_1=16^3 32$, $V_2=24^3 48$, the physical volumes
being $(1.5\fm)^3$ and $(2.3\fm)^3$, respectively. The parameter $\rho$ was 
set to $1.4$.
Our quark masses are
$m_q~=~0.028, 0.056, 0.084\ {\rm and}\ 0.14$, which corresponds to pion masses
ranging from $440-950 \MeV$. On the $24^348$ lattice we have simulated two more
masses down to $m_\pi \approx 300 \MeV$, but with our current statistics the errors
on these data points are too large to draw any conclusions from them.

 In order to remove $O(a)$ errors from the
three-point functions, we employ the method of \cite{capitani1},
which amounts to replacing propagators $D^{-1}\Psi$ by
$\frac{1}{1-\frac{m}{2\rho}}D^{-1}\Psi-\frac{1}{2(1-\frac{m}{2 \rho})}\Psi$. 
Jacobi smeared point sources \cite{goeckeler1} with parameters \mbox{$N_s=50$} 
and \mbox{$\kappa_s=0.21$}
have been used in order to obtain a good overlap with the ground state.

The computation of matrix elements follows the procedure outlined in
\cite{goeckeler1}:
We form the ratio
\eq
R=\frac{\langle N(t_{sink})O(\tau)\bar{N}(t_{source}) \rangle}
{\langle N(t_{sink})\bar{N}(t_{source})\rangle}
\label{eq_ratio}
\ee
from which the matrix element can be extracted in the region
$t_{source} < \tau < t_{sink}$. We always set $t_{source}=0$ and
$t_{sink}=13$ (in lattice units), which corresponds to a
distance between source and sink of $1.2 \fm$. So far we have analysed 250 (40)
configurations on the $16^332$ ($24^348$) lattice.

\section{RENORMALIZATION}

\begin{figure}[t]
\begin{center}
\epsfig{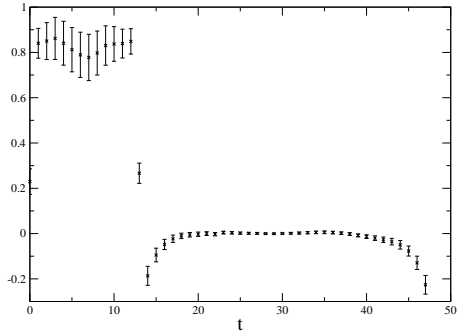}
\vspace*{-0.5cm}
\caption{The ratio of three-point function and two-point function versus
$\tau$ (cf. eq. (\ref{eq_ratio})) for $g_A$ ($m_q=0.028$).}
\label{plot_gA_plat}
\end{center}
\vspace*{-1cm}
\end{figure}

The operators appearing inside the three-point functions have to be
renormalized. For $g_A$, the operator to be used is the axial current $A_\mu$,
the renormalization of which is particularly simple because it can be
obtained from a Ward identity~\cite{giusti2} as
\eq
Z_A=\lim_{t \rightarrow \infty}\frac{2m_q}{m_\pi}
\frac{\langle P(t)P(0)\rangle}{\langle A_4(t)P(0)\rangle}
\label{eq_ZA}
\ee

For overlap fermions, $Z_A$ should agree with $Z_V$ in the chiral limit.
We can compute $Z_V$, making use of current conservation, from the nucleon 
matrix element:
\eq
Z_V \langle N|V_\mu|N\rangle=1
\label{eq_ZV}
\ee

The comparison of $Z_A$ obtained from eq. (\ref{eq_ZA}) to $Z_V$ obtained 
from eq. (\ref{eq_ZV}) is shown in fig. \ref{plot_ZAZV}; extrapolating 
linearly to the chiral limit,
we obtain $Z_A=1.416(20)$, $Z_V=1.426(7)$.

The renormalization of the operators used for the first moment of
the structure function is more difficult; at present, we do not have
a nonperturbative calculation, so we use the value obtained in
tadpole-improved perturbation theory $Z_{v_{2b}}=1.4566$.

\section{RESULTS}

In fig. \ref{plot_gA_plat} we show the ratio of three-point function
to two-point function for the case of $g_A$. One can observe a reasonable
plateau in the region $t_{source} < \tau < t_{sink}$.

\begin{figure}[t]
\begin{center}
\epsfig{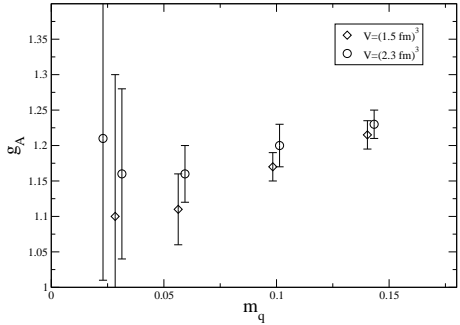}
\vspace*{-0.5cm}
\caption{The nucleon axial charge $g_A$ as a function of the quark
mass.}
\label{plot_gA}
\end{center}
\vspace*{-1cm}
\end{figure}

We show our results for $g_A$ in fig. \ref{plot_gA}.
 A linear extrapolation to the
chiral limit yields $g_A=1.05(5)$ for $V_1=(1.5\fm)^3$
and $g_A=1.13(5)$ for $V_2=(2.3\fm)^3$, which is somewhat lower
than the experimental value $g_A=1.27$.

\begin{figure}[t]
\begin{center}
\epsfig{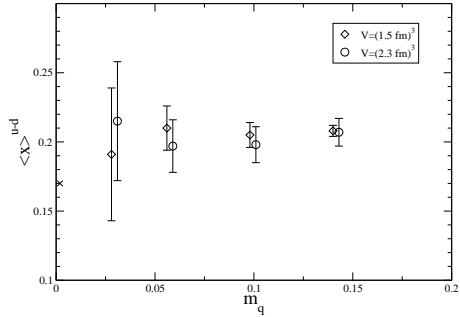}
\vspace*{-0.5cm}
\caption{$\langle x \rangle^{u-d}$ as a function of the quark mass.}
\label{plot_v2b}
\end{center}
\vspace*{-1cm}
\end{figure}

The result for $\langle x\rangle^{u-d}$ is plotted in fig. \ref{plot_v2b}. There appears to be almost no
dependence on the quark mass or on the lattice volume.
In the chiral limit, we obtain $\langle x \rangle^{\overline{MS}}(2\GeV)=0.20(2)$, which is
closer to the phenomenological value than previous results obtained with
Wilson-like fermions. One must bear in mind however that we are still
lacking a nonperturbative value for the renormalization constant.

\section{CONCLUSIONS}

We have determined $g_A$ and $\langle x \rangle^{u-d}$ from quenched overlap fermions.
For $g_A$, our results are in agreement with most previous determinations:
there is some dependence on the lattice volume; on our larger volume
($V=(2.3\fm)^2$ we find $g_A=1.13(5)$.

For $\langle x \rangle^{u-d}$ we see only a weak dependence on volume and quark mass;
the result is considerably closer to the phenomenological value
than most previous lattice results; however, a large systematic uncertainty
remains as long as we do not have a nonperturbative value for the
renormalization constant.


\section*{ACKNOWLEDGEMENTS}

The numerical calculations were performed at the HLRN
(IBM pSeries 690) and at NIC J\"ulich (IBM pSeries 690).
We thank these institutions for their support.



\end{document}